\documentclass[copyright,creativecommons]{eptcs}
\providecommand{\event}{GASCom 2026} 
\usepackage[basic]{complexity}
\usepackage{latexsym}
\usepackage{stmaryrd}
\usepackage{amssymb}
\usepackage[dvips]{graphicx}
\usepackage{amsmath,amsopn,xcolor,xspace,cancel,nccmath}
\usepackage{iftex}

\ifpdf
  \usepackage{underscore}         
  \usepackage[T1]{fontenc}        
\else
  \usepackage{breakurl}           
\fi

\title{On the Counting Sequence of Z-convex Polyominoes}
\author{Luca Castelli\qquad\qquad Paolo Massazza
\institute{Department of Theoretical and Applied Sciences\\
University of Insubria\\Varese, Italy}
\email{lcastelli@studenti.uninsubria.it 
\qquad\qquad  paolo.massazza@uninsubria.it}
}
\def\titlerunning{Z-convex polyominoes}
\def\authorrunning{L. Castelli \& P. Massazza }

\def\high(#1){\ensuremath{\text{high}\!\left(#1\right)}\xspace}
\def\low(#1){\ensuremath{\text{low}\!\left(#1\right)}\xspace}
\def\last(#1){\ensuremath{\textsc{last}\,\!\!\left(#1\right)}\xspace}
\def\first(#1){\ensuremath{\textsc{first}\,\!\!\left(#1\right)}\xspace}
\def\Area(#1){\ensuremath{\text{A}\,\!\!\left(#1\right)}\xspace}
\newtheorem{definition}{Definition}
\DeclareMathOperator{\olap}{\rotatebox[origin=c]{90}{$\rightleftarrows$}}
\begin{document}

\newclass{\A}{A}
\newclass{\ZConv}{ZConv}
\newclass{\AConv}{AConv}
\newclass{\DConv}{DConv}
\newclass{\LR}{LR}
\newclass{\LCR}{LCR}
\newclass{\LFC}{LFC}
\newclass{\LFCR}{LFCR}
\newclass{\LC}{LC}
\newclass{\T}{T}
\newclass{\SSS}{S}
\newclass{\F}{F}
\newclass{\CC}{C}
\newclass{\ZZ}{Z}
\newcommand{\Z}{\ensuremath{\mathbb{Z}}\xspace}
\newcommand{\ignore}[1]{}
\newcommand\mathstack[2]{\genfrac{}{}{0pt}{}{#1}{#2}}
\maketitle
\newcommand{\ie}{\emph{i.e.}\@\xspace}
\begin{abstract}
The degree of convexity of a convex polyomino P is the
smallest integer k such that any two cells of P can be joined by a
monotone path inside P with at most k changes of direction. In this
paper we present a set of formulas and equations that are the basis of a C++ program that allows you to compute the longest counting sequence  known to date (with respect to the area) of convex polyominoes of degree of convexity at most 2.
\end{abstract}

\section{Introduction}
A polyomino is a geometrical figure consisting of a finite set of connected unitary squares (called cells) in the plane
$\Z\times\Z$,
considered up to translations. 
The problem  of counting polyominoes with $n$ cells (\ie \,  of area $n$) is probably one of the fundamental open problems in combinatorial geometry (see problem 37 in \cite{OPP}). Recently, the problem has been solved up to
$n\leq 70$
\cite{BaSh24}.
Due to the difficulty of the problem, simpler classes of polyominoes have been introduced and  studied. In particular, the class of convex polyominoes (polyominoes where the intersection with an infinite horizontal or vertical stripe is a finite segment) and some of its subclasses have been thoroughly investigated  \cite{BM92,BM96,DLNPR04,CR05}.
Sometimes the generating function with respect to the area
$\phi_C(x)=\sum_{n\geq 0}c_nx^n$
of a class $C$ of polyominoes is known, see for instance \cite{KGB17}.
This often allows one to get an estimate of the asymptotic growth of the number of polyominoes of area $n$ in $C$ using standard analytical methods.

When
$\phi_C(x)$
is not known (neither explicitly nor implicitly)
an algorithm for the exhaustive generation of $C$ may produce $c_n$ for small (but still significant) values of $n$. For instance, Constant Amortized Time (CAT) algorithms for generating several classes of polyominoes have been  developed, where the exhaustive generation is done by semiperimeter \cite{DLFR03,DLDFR04} or by area \cite{FoMa17,FoMa18}.

In this paper we consider the class $\ZConv$ of Z-convex  polyominoes. This class has been introduced in \cite{DuRiSc08} and studied in \cite{TV11}. It contains all convex polyominoes $P$ with the property that any two cells of $P$ can be joined by a path in $P$ with at most two changes of direction.
Recently, a combinatorial approach \cite{GuMa24} has been used to prove that Z-convex polyominoes can be counted with respect to the area in polynomial time. By slightly changing some of the combinatorial decompositions in \cite{GuMa24}, we have obtained a new set of recurrence equations and formulas that can be implemented more efficiently. We developed a C++ program that allowed to compute $|\ZConv(n)|$ up to $n=75$.

\section{Basic Notions}
\begin{figure}
    \centering
\includegraphics[width=9.5cm]{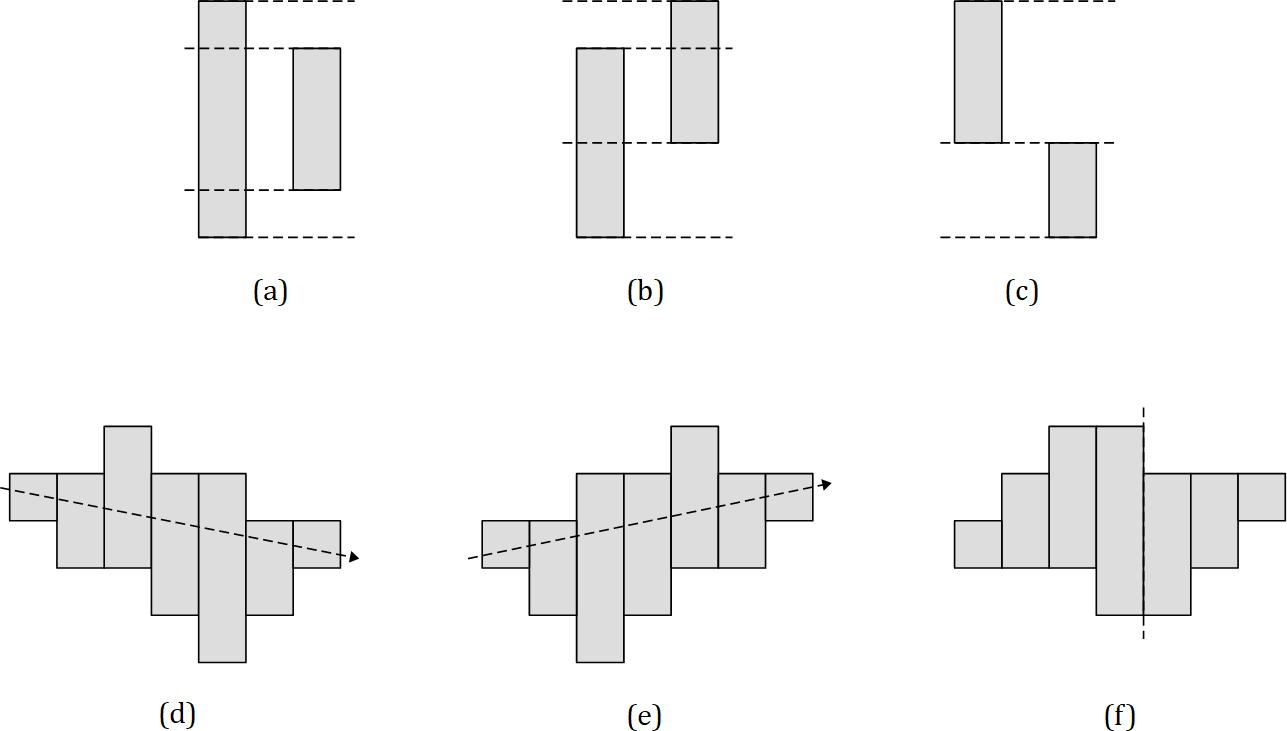}
    \caption{An included column (a), two overlapping columns  (b), two disjoint columns (c), a descending polyomino (d), an ascending polyomino (e) and a polyomino in $\LR$ (f).}\label{fig:notation}
\end{figure}

Let $P$ be a polyomino with an
$r\times c$
minimal bounding rectangle.
We number the columns (resp., rows) of $P$ from left to right (resp., from bottom to top).
The first (resp., last) column of $P$ is denoted by $\first(P)$
(resp., $\last(P)$).
Lastly, we indicate by $\Area(P)$ the \emph{area} of $P$, that is, the number of its cells.
We consider a $Z$-convex polyomino as the result of the concatenation of polyominoes belonging to well-known subclasses of convex polyominoes (stacks $\S$, rectangles $\T$, Ferrers diagrams $\F$, parallelograms $\C$).
We denote by $\low(j)$ (resp., $\high(j)$) the row index of the bottom cell (resp., top cell) of column $j$.
We introduce some binary relations on the set of columns of a convex polyomino, which play a special role in the decomposition of a Z-convex polyomino.
\begin{definition}
Let $i$ and $j$ be two columns of a convex polyomino $P$. We say that
\begin{itemize}
    \item  $i$ \emph{includes}  $j$, denoted by
$j\subseteq i$,
 if and only if $\low(i) \leq \low(j)$ and $\high(i) \geq \high(j)$,
 see Fig. \ref{fig:notation} (a);
    \item  $i$  and $j$ are \emph{overlapping}, denoted by
    $i\olap j$,
    if and only if $\low(j) < \low(i) \leq \high(j) < \high(i)$ or $\low(i) < \low(j) \leq
    \high(i) < \high(j)$, see Fig. \ref{fig:notation} (b);
    \item  $i$ and $j$ are \emph{disjoint}, denoted by
    $i\asymp j$,
    if and only if $\low(i) > \high(j)$ or $\low(j) > \high(i)$, see Fig. \ref{fig:notation} (c).
\end{itemize}
\end{definition}
In the following, an index $^\bullet$  (resp., $^\circ$) to the right of a class indicates that all polyominoes in the class have (resp., do not have) disjoint columns. Thus, for any class $\A$   we have the partition
$\A=\A^\bullet\cup\A^\circ$.

We define the
\emph{center} $\mathcal{C}(P)$
of a convex polyomino $P$ as the rightmost column $e$ of $P$ such that
$c\subseteq e$
for all columns $c$ to the left of $e$. We also define the \emph{pivot} $\mathcal{P}(P)$  of $P$ as the leftmost column $j$ of $P$ such that
$j\olap \mathcal{C}(P)$
(possibly $\mathcal{P}(P)$
does not exist, that is, $\mathcal{P}(P)=\epsilon$).
If $\mathcal{P}(P)\neq\epsilon$
we say that $P$ is \emph{descending} (resp., \emph{ascending}) if 
$\low(\mathcal{C}(P))>\low(\mathcal{P}(P))$
(resp.,
$\low(\mathcal{C}(P))<\low(\mathcal{P}(P))$),
see Fig. \ref{fig:notation} (d) (resp., (e)).
The set of descending (resp., ascending) convex polyominoes is indicated by $\DConv$ (resp., $\AConv$).
If
$\mathcal{P}(P)=\epsilon$
then $\mathcal{C}(P)$ includes all columns of $P$. This means that
$P$ is in $\T\cup\F\cup\SSS\cup\LR$
where  $\LR$ is the class containing all convex polyominoes that are the concatenation of two polyominoes, 
$P=P_1\cdot P_2$,
with
$P_1\in \SSS\cup\F$,
$P_2\in\T\cup\S\cup\F$
and
$\first(P_2)\subsetneq\last(P_1)$.
 We indicate by
$\DConv_2$
(resp., $\AConv_2$)
the set of descending (resp., ascending) polyominoes of degree of convexity 2.
Thus, one has
$$
\ZConv=\T\cup\SSS\cup\F\cup\LR\cup\AConv_2\cup\DConv_2
$$
where the unions are disjoint. The counting problem for $\ZConv$ is therefore reduced to computing
$|\DConv_2(n)|$
(by symmetry one has
$|\DConv_2(n)|=|\AConv_2(n)|$), since the counting problem for $\T\cup\SSS\cup\F\cup\LR$ is easily solved in polynomial time (see \cite[Sect. 4]{GuMa24}).
Lastly, we recall \cite[Thm. 1]{CaMa14} that $P\in\ZConv$ if and only if for any two disjoint columns $i$ and $j$ of $P$ there exists a column $k$, with
$i<k<j$,
such that
$i\subsetneq k$
and
$j\subsetneq k$.
This characterization is the basis of the combinatorial decompositions that lead to the formulas for computing
$|\DConv_2(n)|$.

\section{Polyominoes decomposition}
We consider the \emph{standard decomposition} of
$P\in\DConv$
as defined in \cite{GuMa24}:
\begin{definition}\label{def:stddec}[standard decomposition]
A polyomino
$P\in\DConv$
is uniquely decomposed as
$P=L\cdot F\cdot C\cdot R$
for suitable polyominoes
$L\in\SSS\cup\T\cup\F$,
$F\in \F\cup \T\cup\{\epsilon\}$,
$C\in\C\cup\T\cup\F$,
and
$R\in\SSS\cup\T\cup\F\cup\{\epsilon\}$
with
$\last(L)=\mathcal{C}(P)$,
$\first(C)=\mathcal{P}(P)$
and
$\low({\last(C)})<\low({\first(R)})$.
\end{definition}
Note that any column $e$ of $F$ (if $F\neq \epsilon$) satisfies 
$\low({\last(L)})=\low(e)$
and that
$\last(F)\olap\first(C)$
or
$\last(F)\subsetneq\first(C)$.
Furthermore, 
if $R\neq\epsilon$
one has
$\first(R)\subsetneq\last(C)$.
We point out that by \cite[Thm. 1]{CaMa14},
given
$P\in\DConv_2$
and its standard decomposition
$P=L\cdot F\cdot C\cdot R$,
 each column $c$ of $P$ to the right of $\last(L)$ satisfies the relation
$c\olap\last(L)$
or
$c\subsetneq\last(L)$.

\begin{figure}
    \centering
\includegraphics[width=9cm]{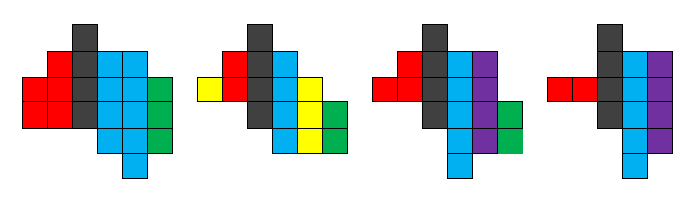}
    \caption{From left to right: $P\in\ZZ_2^\circ\setminus\mbox{u-}\ZZ_2^\circ$,
    $P\in\ZZ_2^\bullet\setminus\mbox{t-}\ZZ_2^\bullet$,
    $P\in\mbox{t-}\ZZ_2^\bullet$ and 
    $P\in\mbox{u-}\ZZ_2^\circ$.}
    \label{polclasses}
\end{figure}

We indicate by $\LCR_2$ (resp., $\LC_2$, $\LFCR_2$, $\LFC_2$) the subset of $\DConv_2$ containing polyominoes whose standard decomposition is
$L \cdot C\cdot R$
(resp., $L \cdot C$, 
$L \cdot F\cdot C\cdot R$, $L \cdot F\cdot C$).
Polyominoes in $\DConv$  will be recursively decomposed into polyominoes belonging to three particular subclasses, see Fig. \ref{polclasses}. The first one is the class $\ZZ_2$ (slightly modified compared to that defined in
\cite{GuMa24}).
\begin{definition}[$\ZZ_2$]\label{def:Z2}
$\ZZ_2$ is the set of all $P$ in $\LCR_2\cup\LC_2$
such that: 1)
for all columns $l$ to the left of $\mathcal{C}(P)$,
one has
$\Area(l)<\Area({\mathcal{C}(P)})$
and
$l\subsetneq \mathcal{P}(P)$;
2)
    $c\olap \mathcal{C}(P)$
    for all columns $c$ of $R$
    (if $R\neq\epsilon$).
\end{definition}
The following two classes are new with respect to \cite{GuMa24} and are introduced to obtain simpler equations.
\begin{definition}[$\mbox{t-}\ZZ_2^\bullet$]\label{def:TZ2}
The set
$\mbox{t-}\ZZ_2^\bullet$
contains all
$P\in\ZZ_2^\bullet$
that can be written as
$P=L\cdot C\cdot E\cdot R$
 for suitable
$L\in\SSS\cup\F$,
$C\in\C\cup\T\cup\F$,
$E\in\SSS\cup\T\cup\F\cup\{\epsilon\}$,
$R\in\SSS\cup\T\cup\F$
 such that:
 \begin{itemize}
 \item $\last(L)=\mathcal{C}(P)$,
 $\first(C)=\mathcal{P}(P)$
 and
 $\low({\last(C)})<\low({\first(E)})$ (if $E\neq \epsilon$);
    \item
    $L$ has at least two columns and
    $l\subsetneq \last(C\cdot E)$
    for all columns $l$ to the left of
    $\mathcal{C}(P)$;
    \item
    $\low({\last(C\cdot E)})\leq\low({\first(R)})$
    and
    $\first(L)\asymp\first(R)$.
\end{itemize}
\end{definition}

\begin{definition}[$\mbox{u-}\ZZ_2^\circ$]\label{def:UZ2}
The set
$\mbox{u-}\ZZ_2^\circ$
contains all
$P\in\ZZ_2^\circ$
that can be written as
$P=L\cdot C\cdot E$
for suitable
$L\in\SSS\cup\F$,
$C\in\C\cup\T\cup\F$
and
$E\in\SSS\cup\T\cup\F\cup\{\epsilon\}$
 such that:
 \begin{itemize}
    \item $\last(L)=\mathcal{C}(P)$,
    $\first(C)=\mathcal{P}(P)$
    and
    $\low({\last(C)})<\low({\first(E)})$ (if $E\neq \epsilon$);
    \item
    $L$ has at least two columns and
    $l\subsetneq \last(C\cdot E)$
    for all columns $l$ to the left of
    $\mathcal{C}(P)$.
\end{itemize}
\end{definition}
\subsection{Computing $|\DConv_2(n)|$}
\begin{figure}
    \centering
\includegraphics[width=6cm]{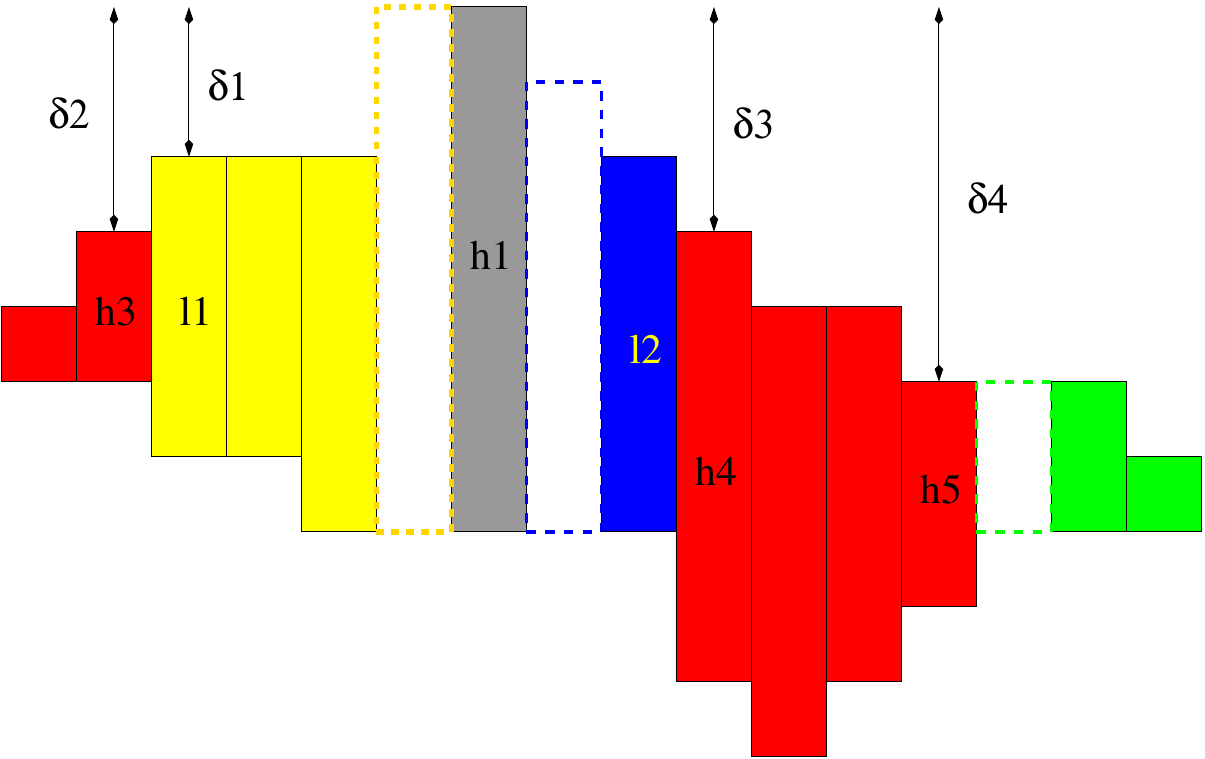}
\caption{Computing $|\DConv_2(n)|$: dashed columns are (artificial) extra-columns, see formula (\ref{eq:DconvCount}). The grey column and the red columns form the polyomino counted by $Z_2(a_1,h_1,h_3,h_4,h_5,\delta_2,\delta_3,\delta_4)$, blue columns form a polyomino in $\F\cup\T$ (counted by $S(a_3+h_1-1,h_1-1,l_2,0)$), green columns form the polyomino in $\S\cup\F\cup\T$ (counted by $S(n-a_1-a_2-a_3+h_1-\delta_4,h_1-\delta_4)$)
and the yellow columns form the polyomino counted by $S(a_2+h_1,h_1,l_1,\delta_1)$.}\label{fig:Dconv}
\end{figure}

\begin{figure}
    \centering
\includegraphics[width=4.4cm]{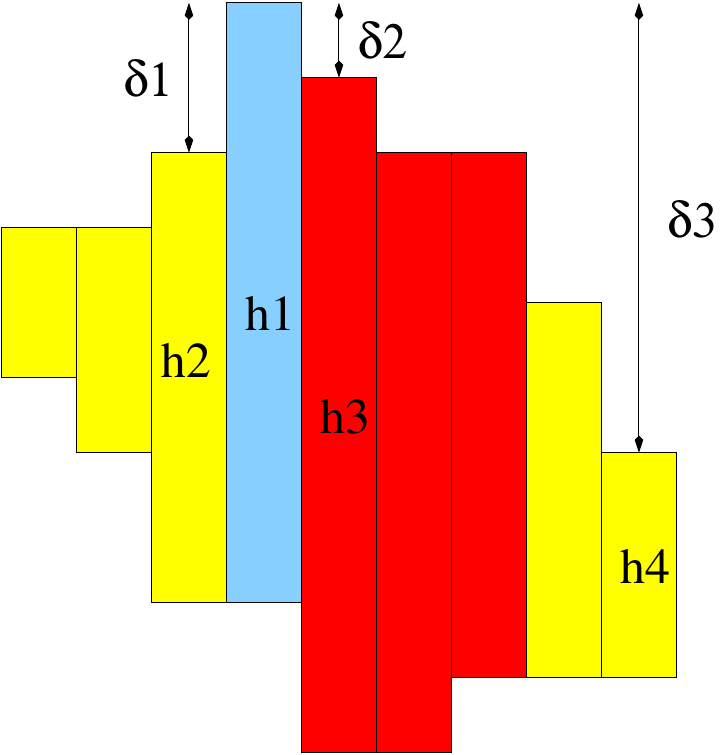}
\caption{A polyomino counted by $Z_2(n,h_1,h_2,h_3,h_4,\delta_1,\delta_2,\delta_3)$.}\label{fig:zpol}
\end{figure}

The first step is to refine the standard decomposition of $P\in\DConv_2$ in order to obtain
$P'\in\ZZ_2$
and some other simpler polyominoes.
Let
$P=L\cdot F\cdot \mathcal{P}(P)\cdot C\cdot R$.
We split $L$ into three parts,
$L=L_1\cdot L_2\cdot\mathcal{C}(P)$,
where
$\last(L_1)\subsetneq\mathcal{P}(P)$
and
$\first(L_2)\olap\mathcal{P}(P)$.
Then, we split
$R$ into two parts,
$R=R_1\cdot R_2$,
with
$\last(R_1)\olap\mathcal{C}(P)$
and
$\first(R_2)\subsetneq\mathcal{C}(P)$.
By construction it follows that
$L_1\cdot \mathcal{C}(P)\cdot\mathcal{P}(P)\cdot C\cdot R_1$
belongs to $\ZZ_2$
(see the red columns and the grey column in Fig. \ref{fig:Dconv}).
\ignore{
We denote by
$DC_2(n,h)$
the number of $P\in\DConv_2(n)$
such that
$\Area(\mathcal{C}(P))=h$
and
$\Area(l)<h$
for all columns $l$ to the left of $\mathcal{P}(P)$.
$$|\DConv_2(n)|=\sum_{h=2}^{n-2}\sum_{e=0}^{\lfloor(n-2)/h\rfloor-1}DC_2(n-e\cdot h,h).$$
}
This leads to a formula for $|\DConv_2(n)|$,
\ignore{
We briefly recall the function $\oplus$ formally defined in \cite{GuMa24}. The idea is that the standard decomposition of any $P\in\DConv_2$ can be refined to obtain one component in $\ZZ_2$ and other  two components, the former (optional) in $\F$, the latter in $\T\cup\SSS\cup\F\cup\LR$. Thus, one has
$\oplus:(\F\cup\{\epsilon\})\times (\T\cup\SSS\cup\F\cup\LR)\times\ZZ_2\mapsto \DConv_2\cup\{\bot\}$,
where
$\oplus(A,B,C)$
is defined (see Fig. \ref{oplus}) if and only if
$\mathcal{P}(B)=\mathcal{P}(C)$,
$B$ and $C$ have the same number of columns equal to 
$\mathcal{P}(B)$ to the left of $\mathcal{P}(B)$,
$\first(B)\olap\mathcal{C}(C)$,
$\Area({\first(A)})<\Area({\mathcal{P}(B)})$
and
$\Area({\last(A)})\geq k$,
where $k$ is the number of cells in $\mathcal{P}(C)$ that have a cell to the right. As a matter of fact one has
\begin{figure}
    \centering
\includegraphics[width=9cm]{oplus.png}
    \caption{The $\oplus$ function.}
    \label{oplus}
\end{figure}
\begin{equation}\label{mainEQ}
    \DConv_2  =  \bigcup_{{F\in\F\cup\{\epsilon\}},{{P\in\T\cup\SSS\cup\F\cup\LR},{Q\in\ZZ_2}}} \oplus(F,P,Q)
\end{equation}
(the unions are disjoint), and
}
\begin{align}\label{eq:DconvCount}
|\DConv_2(n)|   = &
\sum_{\mathstack{h_1,a_1,h_3,\delta_2}{h_4,\delta_3,h_5,\delta_4}} Z_2(a_1,h_1,h_3,h_4,h_5,\delta_2,\delta_3,\delta_4)\cdot \nonumber \\
&\sum_{a_2,l_1,\delta_1} S(a_2+h_1,h_1,l_1,\delta_1)\cdot\sum_{a_3,l_2} S(a_3+h_1-1,h_1-1,l_2,0)\cdot \nonumber\\
& \cdot S(n-a_1-a_2-a_3+h_1-\delta_4,h_1-\delta_4),
\end{align}
where:
\begin{itemize}
\item
$Z_2(n,h_1,h_2,h_3,h_4,\delta_1,\delta_2,\delta_3)$
is the number of polyominoes $P$ in $\ZZ_2(n)$ whose standard decomposition 
$P=L\cdot C$ or $P=L\cdot C\cdot R$
satisfies the following conditions (see Fig. \ref{fig:zpol}):
\begin{itemize}
\item
$L=L_1\cdot \mathcal{C}(P)$,
$\Area({\mathcal{C}(P)})=h_1$,
$\Area({\last(L_1)})=h_2$
(remark: $h_2<h_1$),
$\Area({\mathcal{P}(P)})=h_3$
and
$\Area({\last(P)})=h_4$;
\item
$\high({\mathcal{C}(P)})-\high({\last(L_1)})=\delta_1$,
$\high({\mathcal{C}(P)})-\high({\mathcal{P}(P)})=\delta_2$ and
$\high({\mathcal{C}(P)})-$ \\$\high({\last(P)})=\delta_3$.
\end{itemize}
\item
$S(n,p,q,i)$ is the number of $P\in\T\cup\SSS\cup \F$ of area $n$, with largest (resp., smallest) column of area $p$ (resp., $q$) and such that
$|\low({\first(P)})-\low({\last(P)})|=i$;
\item $S(n,p)$ is the number of $P\in\T\cup\SSS\cup \F$ of area $n$, with largest column of area $p$.
\end{itemize}
Notice the trick of adding (artificial) extra-columns of area $h_1$, $h_1-1$ and $h_1-\delta_4$
in formula (\ref{eq:DconvCount}), to deal with the cases $L_2=\epsilon$ or $F=\epsilon$ or $R_2=\epsilon$.
In the sequel we indicate by 
$C(n,h_1,h_2,i)$
the number of polyominoes $P$ of area $n$ in
$\C\cup\T\cup\F$
with
$\Area({\first(P)})=h_1$,
$\Area({\last(P)})=h_2$
and
$\low({\first(P)})-\low({\last(P)})=i$.
In \cite{GuMa24} it is shown how to compute
$C(n,h_1,h_2,i)$
efficiently.
\subsubsection{Computing $Z_2^\bullet(\boldsymbol{\alpha})$}
\ignore{
Since
$\ZZ_2=\ZZ_2^\bullet\cup\ZZ_2^\circ$
and
$Z_2(n,h_1,h_2,h_3,h_4,\delta_1,\delta_2,\delta_3,e)=Z_2(n-(e-1)\cdot h_1,h_1,h_2,h_3,h_4,\delta_1,\delta_2,\delta_3,1)$
(for $e>1$),
First we focus  on finding a formula for computing $Z_2^\bullet(\boldsymbol{\alpha})$,
where
$\boldsymbol{\alpha}=n,h_1,h_2,h_3,h_4,\delta_1,\delta_2,\delta_3$.
}
We introduce a refinement (see Fig. \ref{shuffle}) of the standard decomposition
$L\cdot C\cdot R$
of a given
$P\in\ZZ_2^\bullet\setminus\mbox{t-}\ZZ_2^\bullet$.

Write $L$ as
$L=L'\cdot D$,
with
$D=\mathcal{C}(P)$,
and let $e$ be the rightmost column of $P$ such that
$\last(L')\subsetneq e$
($e$  exists by Def. \ref{def:Z2}).
Then,
let $c$ be the leftmost column
in $L'$ that is
included in $e$ but not in column
$e+1$.
Lastly, consider the leftmost column $f$ in $C\cdot R$
such that
$f\asymp c$ (possibly $f=\epsilon$).
We stress that if $f$ belongs to $C$ one has
$\low(f')=\low(e)$ for any column $f'$ of $C$ to the right of $e$.
Indeed, if
$\low(f)<\low(e)$
no column of $P$ includes both $f$ and $c$, and so the degree of convexity of $P$ is greater than 2
by \cite[Thm. 1]{CaMa14}.
So, we have five different cases (see Fig. \ref{shuffle}), depending on the component to which the columns $e$ and $f$ belong:
\begin{figure}
    \centering
\includegraphics[width=14.4cm]{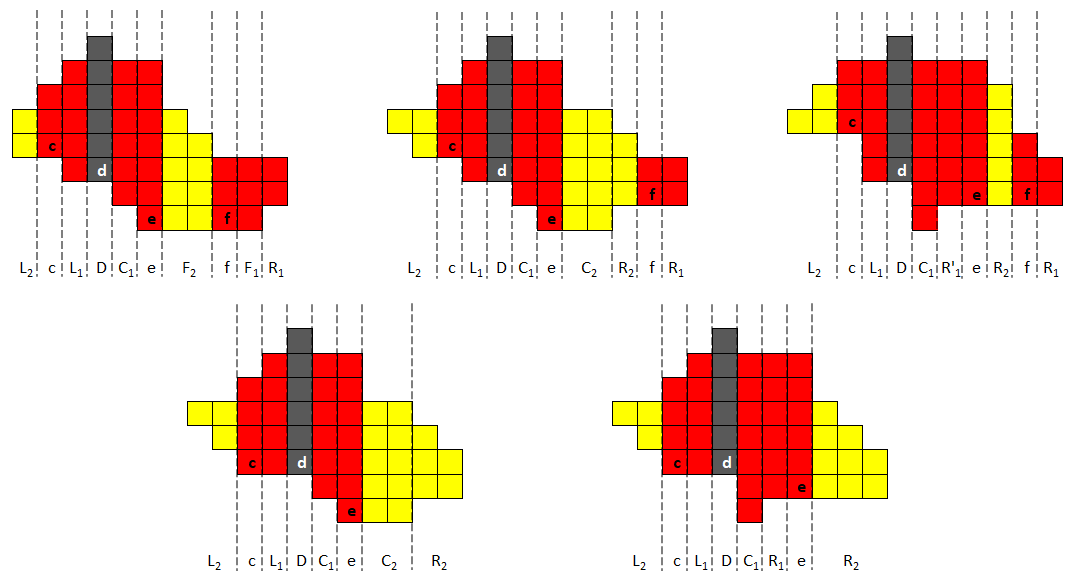}
    \caption{Refining the standard decomposition. Cases 1--5 (from top to bottom and from left to right).}
    \label{shuffle}
\end{figure}
\begin{enumerate}
\item\label{decZC}
$P=L_2\cdot c\cdot L_1\cdot D\cdot C_1\cdot e\cdot F_2\cdot f\cdot F_1\cdot R_1$  (if $e,f\in C$),
\item\label{decZR}
$P=L_2\cdot c \cdot L_1\cdot D\cdot C_1\cdot e \cdot C_2\cdot R_2\cdot f\cdot R_1$ (if $f\in R$ and $e\in C$),
\item
$P=L_2\cdot c \cdot L_1\cdot D\cdot C_1\cdot R'_1\cdot e \cdot R_2\cdot f\cdot R_1$ (if $e,f\in R$),
\item\label{decsZR}
$P=L_2\cdot c \cdot L_1\cdot D\cdot C_1\cdot e \cdot C_2\cdot R_2$ (if $f=\epsilon$ and $e\in C$).
\item\label{decsZRb}
$P=L_2\cdot c \cdot L_1\cdot D\cdot C_1\cdot R_1\cdot e \cdot R_2$ (if $f=\epsilon$ and $e\in R$).
\end{enumerate}
With respect to Fig. \ref{shuffle}, the polyominoes obtained by joining the black column (the center) and the red columns belong to
$\mbox{t-}\ZZ_2^\bullet$
(cases 1--3) or to
$\mbox{u-}\ZZ_2^\circ$
(cases 4--5), whereas the polyominoes obtained by joining the black column and the yellow columns belong to $\ZZ_2$ (cases 1--3) or to $\ZZ_2^\bullet$ (cases 4--5).
\ignore{
We consider the polyominoes 
$P'=c\cdot L_1\cdot D\cdot C_1\cdot e\cdot f\cdot F_1\cdot R_1$
and
$Q'=L_2\cdot D\cdot F_2$
(case 1),
$P''=c\cdot L_1\cdot D\cdot C_1\cdot e\cdot f\cdot R_1$
and
$Q''=L_2\cdot D\cdot C_2\cdot R_2$
(case 2),
and
$P'''=c\cdot L_1\cdot D\cdot C_1\cdot e$
and
$Q'''=L_2\cdot D\cdot C_2\cdot R_2$
(case 4).
By construction it follows that $P'$ and $P''$
belong to
$\mbox{t-}\ZZ_2^\bullet$,
whereas
$P'''$
belongs to
$\mbox{u-}\ZZ_2^\circ$.
We also have that
$Q'$
and
$Q''$
are in
$\ZZ_2$,
whereas $Q'''$ is in
$\ZZ_2^\bullet$.
}
Let
$\boldsymbol{\alpha}=n,h_1,h_2,h_3,h_4,\delta_1,\delta_2,\delta_e$.
The idea we exploit
is that a polyomino $P$ counted by
$Z_2^\bullet(\boldsymbol{\alpha})$
is uniquely obtained by piecewise combining two polyominoes $A$ and $B$ that are determined by the decomposition of $P$ seen above.
In other words, one has
\begin{align}
Z_2^\bullet(\boldsymbol{\alpha}) =
\mbox{t-}Z_2^\bullet(\boldsymbol{\alpha})+
\mathcal{A}(\boldsymbol{\alpha})+
\mathcal{B}(\boldsymbol{\alpha})+
\mathcal{C}(\boldsymbol{\alpha})+
\mathcal{D}(\boldsymbol{\alpha}),\nonumber
\end{align}
where:
\begin{description}
\item[$\mathcal{A}(\boldsymbol{\alpha})$] counts all polyominoes with decomposition (3), or decomposition (2) and $C_2=\epsilon$;
\item[$\mathcal{B}(\boldsymbol{\alpha})$]
counts all polyominoes with decomposition (1), or decomposition (2) and $C_2\neq\epsilon$;
\item[$\mathcal{C}(\boldsymbol{\alpha})$]
counts all polyominoes with decomposition (5), or decomposition (4) and $C_2=\epsilon$;
\item[$\mathcal{D}(\boldsymbol{\alpha})$]
counts all polyominoes with decomposition (4) and $C_2\neq\epsilon$.
\end{description}

By considering Fig. \ref{fig:z2subclasses} (a) and using (artificial) extra columns of area $x_2-1$ (for the magenta component) and $\kappa=i_1+j_1+x_1-e$ (for the yellow component to the left), one has
\begin{align}
\mathcal{A}(\boldsymbol{\alpha})= & \sum_{a_1,x_1,j_1}S(a_1,h_2,x_1,j_1)\cdot
\sum_{a_2,x_2,j_2} C(a_2,x_2,h_3,j_2)\cdot 
\sum_{a_3,x_3,j_3}
S(a_3+x_2-1,x_2-1,x_3,j_3)\cdot \nonumber\\
\cdot & \sum_{a_4,k_1,e} S(a_4,k_1,h_4,i_3-e)\hspace{2pt}\cdot S(n-h_1-a_1-a_2-a_3-a_4+\kappa,\kappa).\nonumber
\end{align}
Similarly, by considering Fig. \ref{fig:z2subclasses} (b) one has
\begin{align}
\mathcal{B}(\boldsymbol{\alpha})=& \sum_{a_1,x_1,j_1} S(a_1,h_2,x_1,j_1)
\cdot\sum_{a_2,x_2,j_2} C(a_2,x_2,h_3,j_2)\cdot
\sum_{a_4,x_4,j_4} S(a_4,x_4,h_4,j_4)\cdot  \sum_{\mathstack{k_3,e_2,k_4}{e_3,k_2,e_1}}
(Z_2^\bullet(\boldsymbol{\beta})+Z_2^\circ(\boldsymbol{\beta})),\nonumber
\end{align}
where
$\boldsymbol{\beta}=n-a_1-a_2-a_4,h_1,k_2,k_3,k_4,e_1,e_2,e_3$.
With respect to Fig. \ref{fig:z2subclasses} (c), one has
\begin{align}\label{ceq}
\mathcal{C}(\boldsymbol{\alpha})= & \sum_{a_1,x_1,j_1}S(a_1,h_2,x_1,j_1)\cdot
\sum_{a_2,x_2,j_2} C(a_2,x_2,h_3,j_2)
\cdot \sum_{a_3,x_3,j_3}
S(a_3+x_2-1,x_2-1,x_3,j_3)\cdot\nonumber \\
\cdot& \sum_{a_4,k_1,e_1} S(a_4,k_1,h_4,i_3-e_1)\cdot
\sum_{e_2,k_2} S(n-h_1-a_1-a_2-a_3-a_4+\kappa,\kappa,k_2,e_2-e_1),
\end{align}
where
$\kappa=i_1+j_1+x_1-e_1$.
Lastly, from Fig. \ref{fig:z2subclasses} (d) we obtain
\begin{align}
\mathcal{D}(\boldsymbol{\alpha})= & \sum_{a_1,x_1,j_1} S(a_1,h_2,x_1,j_1)\cdot\sum_{a_2,j_2,x_2} C(a_2,x_2,h_3,j_2)\hspace{2pt}\cdot
\sum_{k_3,e_2,e_1,k_2}
Z_2^\bullet(\boldsymbol{\beta}),\nonumber
\end{align}
where
$\boldsymbol{\beta}=n-a_1-a_2,\allowbreak h_1,\allowbreak k_2,\allowbreak k_3,\allowbreak h_4,\allowbreak e_1,\allowbreak e_2,\allowbreak i_3$.

\subsubsection{Computing
$\mbox{t-}Z_2^\bullet(\mathbf{\alpha})$
and $Z_2^\circ(\mathbf{\alpha})$}
Formulas for 
$\mbox{t-}Z_2^\bullet(\mathbf{\alpha})$
and
$Z_2^\circ(\mathbf{\alpha})$
follow directly from their definitions.
Indeed, one has
(see Fig. \ref{fig:z2subclasses} (e)) 
\begin{align} 
\mbox{t-}Z_2^\bullet(\boldsymbol{\alpha}) = & \sum_{a_1,x_1,j_1}S(a_1,h_2,x_1,j_1)\cdot \sum_{a_2,x_2,j_2}C(a_2,x_2,h_3,j_2)\cdot\sum_{a_3,x_3,j_3}S(a_3+x_2-1,x_2-1,x_3,j_3)\cdot
\nonumber \\
\cdot& \hspace{2pt} S(n-h_1-a_1-a_2-a_3+\gamma,\gamma,h_4,i_3-i_1-j_1-x_1),\nonumber
\end{align}
where we use (artificial) extra-columns of area $x_2-1$ and $\gamma$, with
$\gamma=i_2+j_2+j_3+x_3-i_1-j_1-x_1$ (if $a_3=0$) or
$\gamma=i_2+j_2+x_2-i_1-j_1-x_1$ (if $a_3>0$).
Lastly, consider the polyomino in $\ZZ_2^\circ$
of Fig. \ref{fig:z2subclasses} (f).
\begin{figure}
    \centering
    \includegraphics[width=0.8\linewidth]{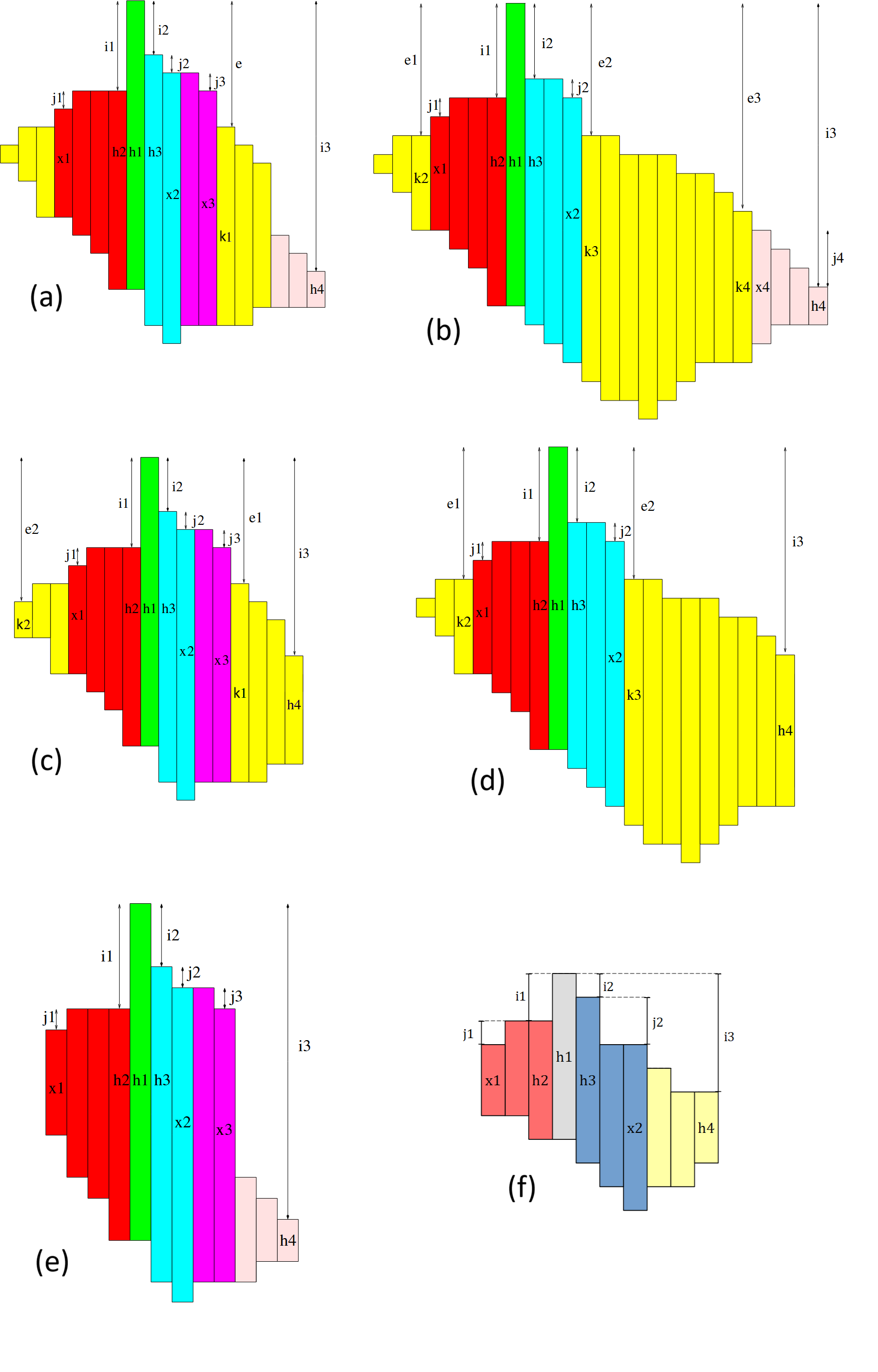}
    \caption{From top to bottom and from left to right:
    a polyomino with decomposition 3 (counted by $\mathcal{A}(\boldsymbol{\alpha})$),
    a polyomino with decomposition 2 and $C_2\neq \epsilon$ (counted by
    $\mathcal{B}(\boldsymbol{\alpha})$ --
    the green column and the yellow columns form $P$ counted by
    $Z_2(\boldsymbol{\beta})$),
    a polyomino with decomposition  5 (counted by $\mathcal{C}(\boldsymbol{\alpha})$),
    a polyomino with decomposition 4 and $C_2\neq\epsilon$ (counted by $\mathcal{D}(\boldsymbol{\alpha})$),
    $P\in\mbox{t-}Z_2^\bullet$
    and
    $P\in Z_2^\circ$  (counted by \ref{eqz2ndalpha}).
    }
    \label{fig:z2subclasses}
\end{figure}

The center $\mathcal{C}(P)$ (of area $h_1$) and the pivot $\mathcal{P}(P)$ (of area $h_3$) are always present, while in general the red component and the yellow component are
optional. Thus, one has
\begin{align}\label{eqz2ndalpha}
Z_2^\circ(\boldsymbol{\alpha}) &= 
C(n-h_1,h_3,h_4,i_3+h_4-i_2-h_3)\hspace{2pt}+ \nonumber\\
&+ \sum_{a_2,x_2,j_2} C(a_2,x_2,h_3,j_2)\cdot
S(n-h_1-a_2+x_2-1,x_2-1,h_4,i_3-i_2-j_2)\hspace{2pt}+ \nonumber \\
&+ \sum_{a_1,x_1,j_1} S(a_1,h_2,x_1,j_1)\cdot\Bigl( C(n-h_1-a_1,h_4,h_3,i_3-i_2)\hspace{2pt}+
\nonumber\\
&+ \sum_{a_2,x_2,j_2} C(a_2,x_2,h_3,j_2)\cdot
S(n-h_1-a_1-a_2+x_2-1,x_2-1,h_4,i_3-i_2-j_2)\Bigr).
\end{align}

\section{Implementation}
We have developed a C++ program that exploits the formulas and the equations presented in the previous section to compute tables for
$Z_2^\bullet(\mathbf{\alpha})$,
$\mbox{t-}Z_2^\bullet(\mathbf{\alpha})$
and
$Z_2^\circ(\mathbf{\alpha})$.
More precisely, we used a dynamic hash table for
$Z_2^\bullet(\boldsymbol{\alpha})$
(resp., $\mbox{t-}Z_2^\bullet(\mathbf{\alpha})$,
$Z_2^\circ(\mathbf{\alpha})$),
with 64 bits keys encoding the eight parameters
$n,h_1,h_2,h_3,h_4,\delta_1,\delta_2$
and $\delta_3$.
Each parameter takes up 8 bits,
hence the program only works for $n<256$. Furthermore, each table contains entries strictly greater than zero. The construction of the table exploits dynamic programming. This means that when you search for a key
$\boldsymbol{\alpha}$
in the table and you do not found it then
$Z_2^\bullet(\boldsymbol{\alpha})=0$.
The size of the hash table for $n=65$ is approximately  $2^{32}$, compared to the size
$2^{48}$ of a table implemented via an 8-dimensional array (with $O(n^8)$ space complexity).
Each entry in a table is computed in polynomial time using one of the formulas in the previous section. The worst case is represented by formula (\ref{ceq}), which leads to a  time complexity of $O(n^{14})$ (due to 14 nested summations). Actually, special care has been taken to identify necessary conditions on $\boldsymbol{\alpha}$ for which
$Z_2^\bullet(\boldsymbol{\alpha})\neq 0$
(so avoiding unnecessary computations), as well as finding tight limits for summation indices.
This allowed to obtain the counting sequence of $\ZConv$ up to $n=75$ in a couple of days of computation. We point out that the sequence indicated in \cite{GuMa24} turned out to be wrong due to a bug in the prototype program used to generate it. By using the new sequence, the conjecture for $|\ZConv(n)|$ \cite{GuMa24}  is corrected in
$C\cdot \exp(\pi\sqrt{17n/6})/n^{3/2})$, for a suitable constant $C$.
Lastly, the sequence in Table 1 does not currently appear in  OEIS \cite{oeis}.

\begin{table}[h!]\label{countsequence}
    \centering \caption{ $|\ZConv(n)|$ for $1\leq n\leq 75$.}
    \begin{tabular}{c}
    \hline
    1, 2, 6, 19, 55, 148, 370, 874, 1966, 4242, 8838, 17851, 35098, 67356, 126518, 233033, 421696, \\
    750780,
    1316916,
    2278259, 
    3891347,
    6567788,
    10962524,
    18108061,
    29619788,
    48004616,
    \\
    77126190,
    122896541,
    94304762,
    304931206,
    475173306,
    735490162,
    1131122763,
     1728912988,
     \\
     2627129510,
     3969544022,
     5965539010,
     8918685922,
     13267244448,
     19641297340,
    28943118312,
    \\
    42459768630,
62020388694,
90215127592,
130699008236,
188612271962,
271160267498,
\\
388410979355,
554388995796,
788571802426,
1117931208744,
1579711035909,
2225197472424,
\\
3124823947072,
4375066863050,
6107738271355,
8502480896054,
11803531168232,
\\
16342160359758,
22566652780440,
31082270916640,
42704425846610,
58529278188804,
\\
80027301158729,
109167041705331,
148578521572114,
201768586748134,
273404214622133,
\\
369684580110248,
498828852782858,
671714659904174,
902712378703847,
1210773577713870,
\\
1620848784573584,
2165731375716902
\ignore{
350797907519, 499444170806, 708639882712, 1002112444338, 1412540714209,\\ 1984808599052, 2780398734144, 3883311845028, 5408022969255, 7510151515584, 10400739110270,\\ 14365314313088, 19789295317410, 27191768575390, 37270314602040, 50960377670716,\\ 69513746774069, 94602105582945, 128453407239846, 174031156719558,235269684178159,\\317382363101408,427264704189028
}
    \\
    \hline
\end{tabular}
\end{table}
\section*{Acknowledgements}
We would like to thank Andrew R. Conway who provided us with the drawings of the Z-convex polyominoes (generated by a smart program and grouped by area). They were extremely useful for debugging our program. We also thank Anthony J. Guttmann for updating the conjecture on $|\ZConv(n)|$ based on the new sequence.
\ignore{
\begin{figure}
    \centering
    \includegraphics[width=0.8\linewidth]{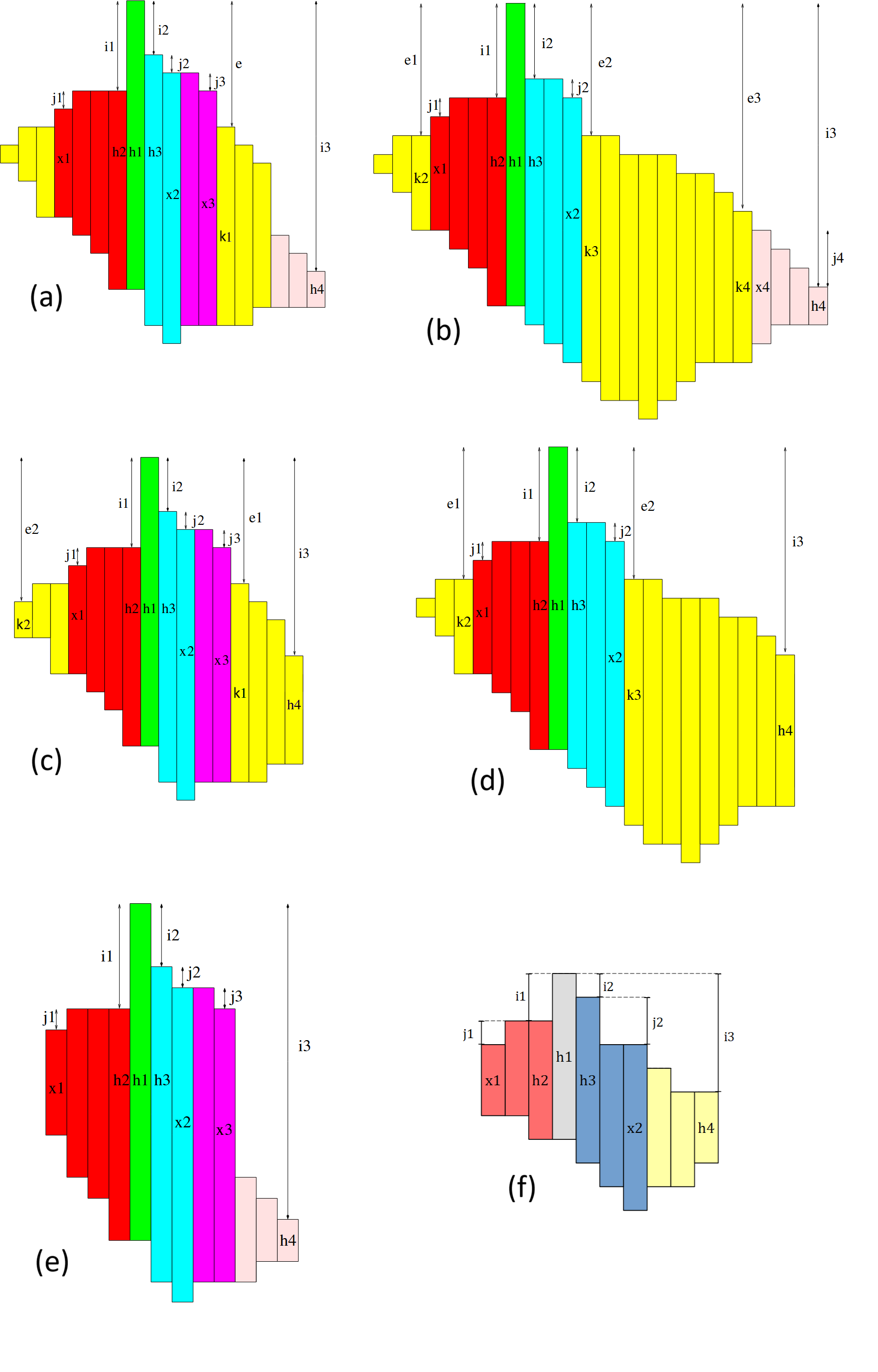}
    \caption{From top to bottom and from left to right:
    a polyomino with decomposition 3 (counted by $\mathcal{A}(\boldsymbol{\alpha})$),
    a polyomino with decomposition 2 and $C_2\neq \epsilon$ (counted by
    $\mathcal{B}(\boldsymbol{\alpha})$ --
    green and yellow columns form $P$ counted by
    $Z_2(\boldsymbol{\beta})$),
    a polyomino with decomposition  5 (counted by $\mathcal{C}(\boldsymbol{\alpha})$),
    a polyomino with decomposition 4 and $C_2\neq\epsilon$ (counted by $\mathcal{D}(\boldsymbol{\alpha})$),
    $P\in\mbox{t-}Z_2^\bullet$
    and
    $P\in Z_2^\circ$  (counted by \ref{eqz2ndalpha}).
    }
    \label{fig:z2subclasses}
\end{figure}
}

\nocite{*}
\bibliographystyle{eptcs}
\bibliography{poli}
\end{document}